# Multi-party Holomeetings: Toward a New Era of Low-Cost Volumetric Holographic Meetings in Virtual Reality

Sergi Fernández[1,2], Mario Montagud[1,3], Gianluca Cernigliaro[1], David Rincón[3]
[1] i2CAT Foundation, Barcelona (Spain)
[2] Universitat Politècnica de Catalunya (UPC). Spain
[3] Universitat de València (UV), Spain
{sergi.fernandez, mario.montagud, gianluca.cernigliaro}@i2cat.net; david.rincon@upc.edu

*Abstract* — Fueled by advances in multi-party communications, increasingly mature immersive technologies being adopted, and the COVID-19 pandemic, a new wave of social virtual reality (VR) platforms have emerged to support socialization, interaction, and collaboration among multiple remote users who are integrated into shared virtual environments. Social VR aims to increase levels of (co-)presence and interaction quality by overcoming the limitations of 2D windowed representations in traditional multi-party video conferencing tools, although most existing solutions rely on 3D avatars to represent users. This article presents a social VR platform that supports real-time volumetric holographic representations of users that are based on point clouds captured by off-the-shelf RGB-D sensors, and it analyzes the platform's potential for conducting interactive holomeetings (i.e., holoconferencing scenarios). This work evaluates such a platform's performance and readiness for conducting meetings with up to four users, and it provides insights into aspects of the user experience when using single-camera and low-cost capture systems in scenarios with both frontal and side viewpoints. Overall, the obtained results confirm the platform's maturity and the potential of holographic communications for conducting interactive multi-party meetings, even when using low-cost systems and single-camera capture systems in scenarios where users are sitting or have a limited translational movement along the X, Y, and Z axes within the 3D virtual environment (commonly known as 3 Degrees of Freedom plus, 3DoF+).

## 1. Introduction.

Advances in information and communications technologies (ICT), particularly in video technologies, have opened the door to the proliferation of real-time multi-user communication tools and services, such as video conferencing. In this context, video conferencing tools have provided society with many benefits like remote meetings, e-learning, and teleworking, thus obviating the need to commute and thereby saving time and costs while reducing the carbon footprint.

Recent years have seen significant research efforts being devoted to not only designing and optimizing multi-party conferencing services (e.g., in terms of latency, media compression, quality, scalability, etc.) [1], but also to better understanding how quality of service (QoS) factors affect the user's perceived quality of experience (QoE) (e.g., [2]). In addition, video conferencing tools have progressively evolved beyond core signaling, encoding and communication aspects and now incorporate added features that support richer interactions and collaboration among remote users (e.g., shared consumption of media, slide presentations, screen sharing) [3]. Many open-source and commercial video conferencing tools now exist and have been widely adopted, like Meet, Teams, and Zoom. All of them have multiple interactive features, and some such as Sceenic[1] even offer co-watching of live content.

In parallel with advances in multi-party communications, the increasing maturity and adoption of immersive technologies has fueled the appearance of a new wave of social virtual reality (social VR) platforms. These social VR platforms support socialization, interaction, and even collaboration among multiple remote users immersed in shared virtual environments [5], thus going beyond the typical mosaics of upper body video representations in traditional video conferencing platforms. Social VR constitutes a promising medium to increase plausibility (i.e., the feeling of realism), presence, immersion, co-presence (i.e., togetherness), quality of communication, naturalness, and comfort, thereby overcoming the current limitations of 2D tiled windowed representations and communications when using traditional multi-party conferencing. Examples of social VR platforms include Mozilla Hubs,[2] Facebook Horizon,[3] and AltspaceVR[4] (by Microsoft, soon to be replaced by the Mesh platform). Most of these platforms rely on using synthetic 3D (cartoonish or human-like) avatars for representing users. In this context, state-of-the-art studies (which are reviewed in Section II) have reported diverse benefits (e.g., in terms of (self-)embodiment, (co-)presence, and quality of communication) not only from using social VR rather than traditional 2D conferencing tools, but also from using video-based or volumetric user representations rather than avatars in social VR.

Despite the fact that social VR represents a promising medium for bringing people together in an immersive, natural, and comfortable manner, the few currently existing studies on this topic have revolved around technological aspects (e.g., [6]) and on shared media consumption experiences (e.g., [5, 7]). Thus, their contributions were focused mostly on the shared VR scenario and content being consumed, although to a lesser extent on the quality of user representations and of user communication when this is the focus, such as in pure multi-user communications like video conferences. A few other studies have focused on analyzing the effectiveness of interaction and collaboration in multi-user social VR scenarios (e.g., [8]), and some even consider aspects of content comprehension in comparison to traditional conferencing solutions (e.g., [9]). However, these later works are restricted to the use of avatar-based social VR solutions.

This paper aims to fill this gap in the literature by using an evolved version of a lightweight and low-cost social VR platform [5, 6], in which remote users are captured by off-the-shelf RGB-D cameras (e.g., Azure Kinect) and are then integrated into shared virtual environments in real-time. Furthermore, we explore the readiness and potential benefits of such a multi-party holographic teleportation (i.e., holoportation) technology to enabler interactive holomeeting and holoconferencing use cases (Figure 1), thus going the medium's originally conceived uses for social co-viewing and entertainment [5, 7].

---

[1] Sceenic, https://www.sceenic.com/ Last accessed in June 2022
[2] Mozilla Hubs, https://hubs.mozilla.com/ Last accessed in June 2022
[3] Facebook Horizon, https://www.oculus.com/facebook-horizon/ Last accessed in June 2022
[4] AltSpaceVR, https://altvr.com/ Last accessed in June 2022

Three key and distinctive aspects from this study can be highlighted. First, unlike previous works that have evaluated user experiences with representations in video-based (i.e., RBG-D) and mesh-based formats [5, 7], this paper focuses on user representations based on point clouds [6], which is a promising volumetric media format for representing natural content like the human body. Second, although the capture sub-system of the presented social VR platform supports both single- and multi-camera setups for obtaining the volumetric point cloud representation, this study uses the former in order to minimize costs, deployment complexity, and computational load while also assessing whether this simpler setup is appropriate for the user experience. Third, unlike previous social VR studies that have paid significant attention to VR storytelling and the associated content representing the shared environment and activities (e.g., collectively watching VR content [5] or a virtual TV show [7]), this study focuses on the user representations and quality of their communications. Specifically, we teleport groups of two and four people to a virtual meeting room around a round table and ask them to play a guessing game in which they make gestures that draw attention to their bodily representations, which allows us to study the naturalness and effectiveness of their interactions. Thus, this paper aims to shed some light on the following research questions (RQ):

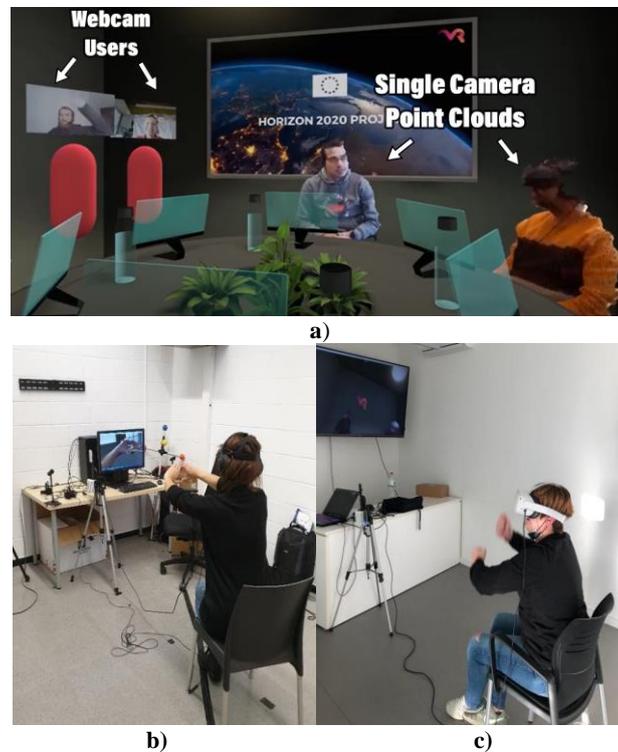

*Figure 1. Holomeeting scenarios using the social VR platform: a) holomeeting with four users, two of them represented as volumetric point clouds; b) and c) evaluation setup, with one RGB+D camera, one desktop or laptop, and one head-mounted display (HMD)*

**RQ1**) Is the current lightweight and low-cost holoportation technology (i.e., the presented social VR platform that provides realistic volumetric user representations) ready and effective enough to conduct interactive and immersive multi-party holoconferencing services?

**RQ2**) What impact does the number of participants have on multi-party holoconferencing services?

**RQ3**) What are the implications of using single-camera capture systems in scenarios where a full volumetric representation is noticeably missing (i.e., four users at fixed positions, equally distributed around a round meeting table, and thus with frontal and side viewpoints)?

On the one hand, this paper reports the results of objective tests on the platform's performance in terms of latency and the usage of computational and bandwidth resources when running the described holomeeting scenarios. This will determine the deployment and usage requirements for such a technology and its associated services. On the other hand, we use diverse questionnaires and semi-structured interviews to conduct subjective tests with $N=32$ participants (i.e., eight groups of four participants, and sixteen groups of two participants) to provide key insights into the limitations, readiness, applicability, and potential of the presented multi-party holoportation technology for conducting holomeetings and related use cases.

Therefore, this paper's contributions offer a valuable resource and output for the scientific community and interested stakeholders, namely in proving the relevance and opportunities provided by this new multi-party communication and interaction medium using holographic representations.

The rest of this paper is structured as follows. Section II briefly reviews the state-of-the-art in traditional multi-party conferencing tools, as well as more recent contributions on the (potential) adoption of social VR and other immersive technologies for real-time multi-user interaction. Section III provides a high-level overview of the end-to-end social VR platform that has been developed and adapted to support the presented use case. Section IV reports on the objective and subjective tests that have been conducted, while also describing the evaluation setup and methodology. Next, Section V provides a discussion about the obtained results and insights. Finally, Section VI presents our conclusions and elaborates on opportunities and the necessity for future work.

## 2. Related Work

This section first provides a brief overview of research contributions on the design, optimization, and evaluation of video conferencing tools, as well as on understanding the impact of QoS factors on perceived QoE. The goal of this first subsection is not to provide an exhaustive review of all research efforts and contributions related to multi-party video conferencing, but to instead highlight key aspects and insights while providing evidence on the topic's relevance and the wide attention it has received in the past three decades.

Finally, this section elaborates on more recent works that recreate immersive meetings by adopting omnidirectional video and 3D environments with integrated users, either as avatar representations (some of them replacing the avatar's head with a squared 2D video window presenting the user's webcam stream) or with realistic volumetric representations.

2.1. *Research on 2D Multi-party conferencing*

Multi-party video conferencing systems and tools have existed for nearly three decades. Since their invention, the research community and industry have devoted major efforts to optimizing their performance in terms of many relevant factors, such as delays (e.g., [10]), scalability (e.g., [11]) and video quality (e.g., [2]), while trying to minimize the consumption of resources and/or maximize the resulting QoE. Likewise, many research works have focused on understanding system design (e.g., [12]), QoS (e.g., [12, 13]) and contextual factors in order to improve the users' perceived QoE (e.g., [14]). In this context, the arrival of recent standard technologies, like Web Real-Time Communication (WebRTC)[5] [15], has represented a huge leap in the development and adoption of web-based multi-party conferencing solutions. In addition, other video conferencing solutions have begun to integrate interactive features like text chat and media sharing to enrich socialization and collaboration (e.g., [16, 17]). The work in [3] reviews the interactive features provided by many of the available web-based multi-party conferencing tools, discusses their potential benefits for social co-viewing, and demonstrates the benefits that these features bring to the user experience. Nowadays, most existing video conferencing tools like Teams, Zoom, and Meet incorporate at least a sub-set of these interactive and collaborative features.

Finally, a more recent study [18] investigated the challenges and requirements for enabling shared 360º video consumption scenarios using head-mounted displays (HMDs) and by incorporating a set of guiding and interaction strategies.

2.2. *Research on Social VR Design Factors and on Immersive Multi-party Meetings*

Compared to shared experiences through 2D screens when using traditional video conferencing tools, social VR aims to deliver experiences that are closer to physical meetings by maximizing the feeling of plausibility, presence, co-presence, as well as the interaction quality (e.g., [5], [8], [9], [19]). Many works have evaluated how the interactions and QoE in social VR are impacted by system design features like avatar formats and realism. The work in [8] suggests that even though social VR systems reduce usability (i.e., they are less easy to use, probably due to lack of familiarity), they increase presence when compared to traditional conferencing systems (even when using 2D screens and avatar-based representations), concluding that both types of tools seem to be effective for conducting online lessons. The work in [20] highlights the potential of VR technologies in effectively supporting interactive conversations and collaborations between remote users, although it reflects on the need for a collaborative design framework addressing key factors and aspects, such as: (i) variability in social immersion; (ii) user and conversational roles; and (iii) effective shared (and spatial) references. The work in [8] reports a longitudinal and exploratory study on individual workload, presence, and emotional recognition in collaborative multi-party virtual environments. The obtained results show that although the reported levels of presence and workload did not vary over time, the adaptation to VR (in terms of interaction with partners and execution of tasks) significantly increased over time, and the reported levels of co-presence were influenced by the task at hand. That study also discusses design implications and suggests future directions for designers and researchers in the field, with a particular focus on how to effectively enable meaningful social interactions and collaboration in VR.

---

[5] WebRTC project page: https://webrtc.org/ Last accessed in June 2022.

Other studies have focused on analyzing the influence and implications of the user's representation format in (social) VR. The work in [21] analyzed factors influencing the sense of embodiment that leads to having a virtual body in immersive environments. In [22], the authors investigated how the avatar's appearance influences the sense of presence in social VR scenarios. By comparing different user representation modalities, the results indicate that even when only the head and hands of motion-controlled avatars were visible, increased feelings of co-presence and behavioral interdependence were produced. The work in [23] investigated the avatar realism, with a focus on realistic body movements to determine the potential influence on social interaction quality, specifically in terms of these realistic movements compensating for missing facial expressions and eye gaze cues. The results indicate that social interactions tend to be impeded by non-realistic avatars, but the absence of important behavioral cues such as gaze and facial expressions can be partially compensated by realistic body movements. The work in [24] investigated communication behavior in embodied VR by comparing the audio-visual communication patterns between two users completing a task under three conditions: (1) face-to-face; (2) social VR with embodied avatars that have an eyebrow ridge and nose, but no other facial features; and (3) social VR with only virtual hands and no visible avatars. The authors concluded that embodied avatars provide high levels of social presence and conversation patterns, being somehow comparable to those in face-to-face interactions. In [25], the effect of avatar realism on self-embodiment and social interactions in social VR was also explored. The results show that realistic avatars are rated significantly more human-like and evoke stronger acceptance in terms of virtual body ownership. Similarly, it was found in [26] that the possibility of personalizing avatars results in an increased sense of body ownership, presence, and dominance.

Beyond using different types of avatars, many works have additionally explored the benefits of adopting video-based user representations in social VR. The work in [27] explored different setups to enable immersive multi-party video conferencing, and it proposed a shared virtual table environment where small groups of users can be integrated via video-based representations. The work in [28] provided initial empirical evidence that, compared with using animated synthetic avatars, real-time and realistic 3D human representations in VR environments result in an increased feeling of presence, as well as physical, emotional, and user state recognition. The work in [29] proposed adopting HMDs in conjunction with single RGB-D cameras (Kinect) for the end-user's capture and representation, concluding that this can also increase engagement and the feeling of immersion while providing an enjoyable embodied telepresence, particularly when compared with traditional 2D social co-viewing tools. The work in [30] presented a video-based social VR platform focused mainly on enabling shared media consumption of stored content. On that platform, users are captured by a single RGB-D camera (Kinect), and the shared VR scenario is represented as a 360º static image. Next, the work in [19] proposed an experimental protocol and a questionnaire for evaluating social VR experiences. By adopting a photo sharing use case, the experiment compared interaction quality, social meaning, and presence levels in three scenarios: (1) face-to-face; (2) Skype; and (3) video-based social VR (using the platform from [30]). The experimental results in that work not only proved that the proposed evaluation methodology was appropriate (i.e., the designed questionnaire was highly reliable), but also that video-based social VR provides an enhanced user experience compared to traditional video conferencing tools like Skype. Based on the promising insights from [19], the work in [5] conducted two additional experiments to assess the benefits of realistic user representations in social VR scenarios. The first experiment recreated a shared video watching scenario in groups of two users under the same three test conditions as in [19]. The obtained results not only reinforced the appropriateness of the evaluation methodology proposed in [19], but they also corroborated that in this video co-viewing use case the benefits also extend to using realistic video-based representations over avatars. Furthermore, that work found that users' experiences and behaviors on this video-based social VR platform were comparable to those in face-to-face scenarios. These results, in turn, encouraged the development of a lightweight and low-cost social VR platform that supports shared 3D environments and real-time captured volumetric representations of users, specifically by using time-varying meshes (TVM) as the representation format, by adopting technological contributions from [31, 32]. Accordingly, the second experiment in [5] consisted of adopting the previously mentioned platform for evaluating a shared co-viewing experience by using a professionally created 3D VR episode as the shared content and not just 2D videos presented on a traditional screen, as in the first experiment. Although different technologies and content stimuli were used in the two experiments, both obtained similar results in terms of presence, togetherness, and interaction quality, which is additional evidence for the potential benefits of using realistic volumetric representations in social VR, regardless of the margins that remain for improvements in interactive volumetric media technology. In [7], the authors studied the use of volumetric user representations in addition to a strategic combination of immersive and traditional media formats for the shared virtual environment. These combinations were applied to a social VR scenario recreating a live TV show, where the users (audience) could interact amongst themselves and with a live presenter while being provided with additional interactive video presentations, obtaining promising results. Finally, the work in [6] further developed the social VR platform in [5] to additionally support volumetric user representations as point clouds, which is a very promising format for representing natural content being under standardization within the Moving Picture Experts Group (MPEG) [33]. The present work relies on an evolved version of this platform to assess the potential of this promising technology for efficiently supporting interactive multi-party holomeetings. To our knowledge, the work presented here provides a groundbreaking assessment and demonstration that real-time point cloud representations are viable and appropriate for conducting interactive multi-party holomeetings, even when using a lightweight platform using low-cost and off-the-shelf equipment.

## 3. Social VR Platform for Holomeetings

This section presents an overview of the novel, lightweight, and hyper-realistic social VR platform that we have further developed (departing from the platforms in [5], [6] and [7]) and adopted for conducting the holomeeting experiments. Figure 2 shows the high-level architecture of the platform, including its key blocks, components, and the information exchanged among them. The architecture is designed to be flexible and modular so that its components and modules can be easily replaced and adapted for testing different variants and configurations. The next subsections describe each of these key components.

### 3.1. *Pipelines for Volumetric Media (Point Clouds)*

#### 3.1.1. Capture & Reconstruction

To enable realistic and fluid volumetric representations of users in the social VR experience, we have integrated a real-time volumetric video capture and reconstruction sub-system based on the work in [6], where participants are represented as point clouds [33]. A point cloud represents a 3D object as an unstructured collection of points with X, Y, and Z coordinates, plus additional attributes that can be, among other things, the RGB color values of an associated point on the volume's surface. Although point clouds are not a complex representation format and therefore do not require heavy pre- and post-processing, they tend to require extensive storage and thus demand huge bandwidth for real-time networked delivery.

The point cloud capture sub-system can adopt setups of 1 to N RGB-D sensors (Figure 2), such as Intel RealSense cameras (e.g., the D415 model[6]) [34] or Azure Kinect[7] [35], which both capture color (RGB) and depth (D) information. In the case that >1 RGB-D sensors are used, they are strategically placed for capturing the full volume, with an effective capture area that has approximately a 1.5m radius. The captured point cloud frames from each sensor are then fused in a common processing station to provide a reconstructed volume as output, as detailed in [5, 6].

The capture module interfaces with the cameras to obtain the RGB and depth images, which are then transformed and converted into a point cloud representation. An intrinsic 4x4 matrix is used per each camera to compute the transformation between RGB and depth images, as well as to create the point cloud for each camera. Likewise, extrinsic matrices are used to convert the per-camera point clouds into world coordinates, which allows for their appropriate fusion into the resulting volumetric point cloud. Further details are provided in [5] and [6].

---

[6] Intel RealSense D415 sensor: https://www.intelrealsense.com/depth-camera-d415/ Last accessed in June 2022.
[7] Azure Kinect: https://azure.microsoft.com/es-es/services/kinect-dk/ Last accessed in June 2022.

Theoretically, there is no limitation in terms of the number of RGB-D sensors that can be used in the capture and reconstruction sub-system. However, it is necessary to consider limitations like physical space, computational resources, hardware connections, and interference between sensors. If a single RGB-D sensor is used, only the portion of volume visible to the sensor will be captured, but this can suffice for certain settings and conditions, as explored in this study.

The point cloud data stream is also provided to the local rendering process in order to enable the user's self-representation in the virtual environment (Figure 2).

3.1.1. Encoding & Transmission

Due to their huge amounts of data, point clouds need to be encoded for real-time distribution over networked scenarios. In recent years, point cloud compression has received significant attention from the scientific community and standardization bodies, like MPEG [33].

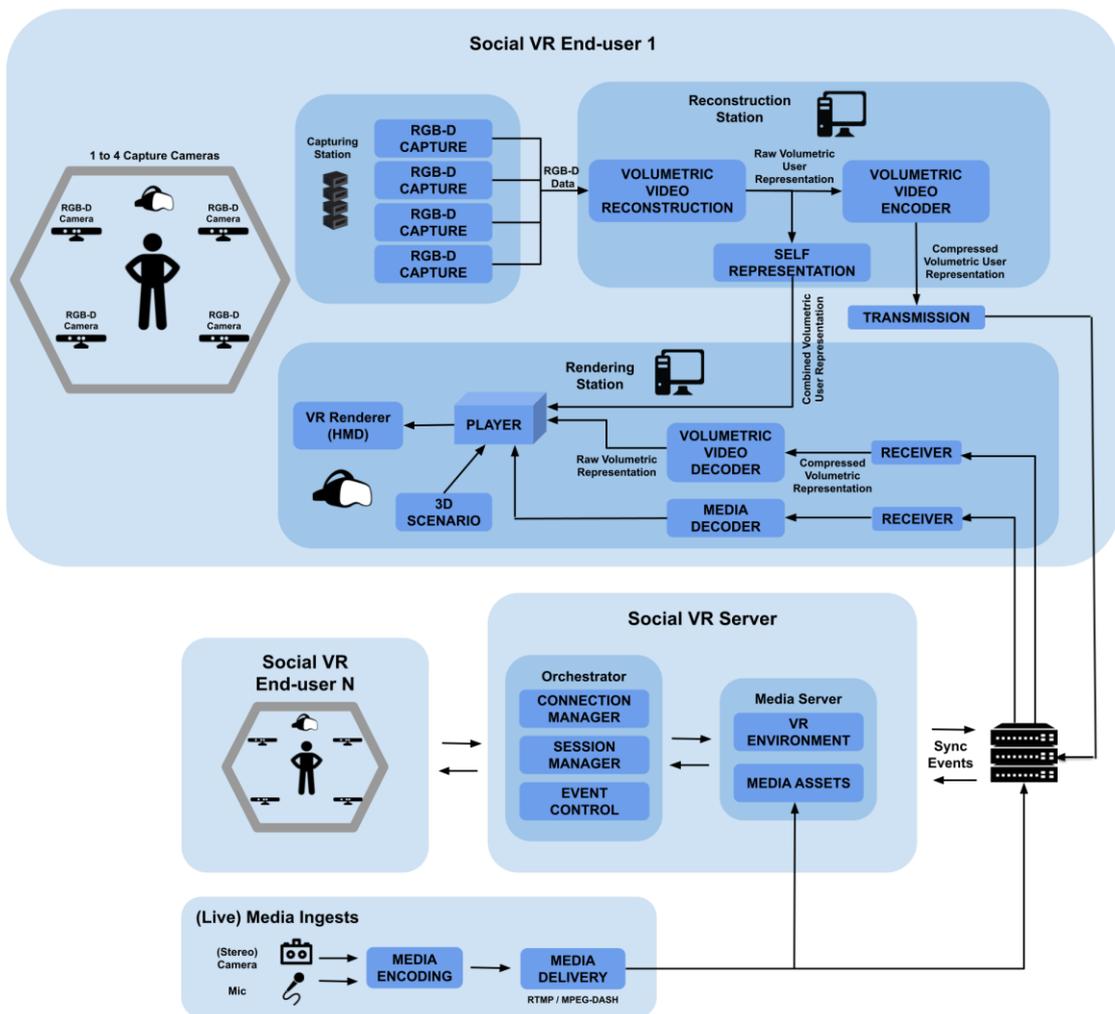

*Figure 2. High-level architecture and flow diagram of the presented social VR platform*

To encode the point clouds, this work has adopted the encoder/decoder from [36, 6]. The codec is based on using intra frames for the whole captured stream; it exploits octree occupancy to represent geometry; and it strategically projects the colors onto a 2D grid using a Joint Photographic Experts Group (JPEG) image compression technique. Such configurations allow for low delay encoding and decoding, thus making it suitable for real-time applications. The adopted implementation of the point cloud encoder allows for encoding up to around 50K points at 15 frames per second (fps). Accordingly, we developed a wrapper to enable smooth real-time encoding of the captured and reconstructed point clouds from the adopted RGB-D sensors.

After compression, the encoded point cloud frames are distributed by means of either Dynamic Adaptive Streaming over HTTP (DASH) [6] or *socket.io*[8] (i.e., socket-based) connections managed by an orchestrator (introduced in Section III.B). Apart from the visual communications channel, the platform integrates a bi-directional audio communication pipeline that also relies on socket connections (*socket.io*) for the data exchange

### 3.2. *Orchestration and Interactive Session Control*

#### 3.2.1. Orchestration

Orchestration components (i.e., orchestrators) are commonly used in video conferencing systems to handle sets of audiovisual and control streams [37, 38]. For the social VR platform presented here, we developed and integrated an orchestrator to deal with session and stream management tasks. The orchestrator handles the remote networking information (e.g., IP addresses, ports, protocols) and performs the following key actions to support multi-party social experiences: i) accommodating all remote users in a shared virtual environment/session; ii) managing the real-time interaction channels and acting as a relay server for media streams; iii) exchanging user positions in the 3D environment; and iv) ensuring a consistent, synchronized experience.

In addition, the orchestrator provides information on potential errors and unexpected behavior in the distributed shared sessions; and it can potentially perform a set of recovery actions in case of connection problems.

#### 3.2.2. Interactive Session Control

Our social VR platform supports the dynamic ingest and control of live and on-demand 2D media sources (Figure 2) to enable interactive social viewing scenarios. For such a purpose, we have developed a pipeline for traditional media formats with the associated capture, retrieval, encoding and delivery components. On the one hand, the pipeline supports the most widely used encoding / decoding formats (e.g., MPEG4-AVC/H.264 [39]), with the preferred settings. On the other hand, the pipeline supports different protocols for media delivery, like Real-Time Messaging Protocol (RTMP), DASH, or even socket.io, which are among the most widely adopted solutions for media streaming. This in turn enables selecting the delivery method that is best suited to the target requirements. For example, RTMP or socket.io may be chosen for low-latency and DASH for scalability and media quality adaptation. A similar pipeline could be adopted for real-time communication purposes, providing support for more lightweight user representations, with an avatar-like body and replacing the head with a 2D window showing the video stream captured by a webcam or any other camera (see Figure 1). That would eliminate the strict need for having an RGB-D sensor available for participating in a holomeeting session with a volumetric representation. However, these pipelines for traditional media formats are not considered in the presented experiment, as its focus is on the volumetric holographic representations and interaction quality when using them. These pipelines and interactive features are just mentioned here for completeness, showing the extra capabilities and flexibility of our platform, without influencing the main scope of the study.

---

[8] Socket.io: https://socket.io/ Last accessed in December 2021.

### 3.3. *Playout*

The final stage for each pipeline is to present the media content on the client side while integrating all social VR interaction modalities.

To this end, we developed a Unity-based player (Windows build) to properly receive, integrate, and present all available streams for the shared VR scenes, the end-user representations, and all other additional assets and media sources that can enrich the experience.

The player includes different components and engines that provide the following features:

- Connection to and interaction with the orchestrator. To communicate with the orchestrator, the user must first log in through the player interface and then create and/or join a shared social VR session by selecting from among a set of desired VR scenarios. During the session, the necessary information will be exchanged to enable interactive and coherent experiences. Finally, at the end of the experience, the session is terminated by freeing up all associated resources.
- Loading or receiving the 3D virtual scenario where the end-users will be teleported. Each user is initially appropriately placed and oriented within the virtual scenario, and later on their positions and orientations are exchanged in order to provide coherent multi-user experiences.
- Receiving the data streams for one's own and others' representations (as point clouds in this work).
- Receiving the data streams from the stored and live media sources, including traditional 2D and stereoscopic 180º/360º video, if activated.
- Seamlessly blending all content formats and streams constituting the social VR experience.
- Ensuring intra-media and inter-media synchronization [40], as well as presenting events in a timely and synchronized manner [41], in coordination with the orchestrator.

The player can run on the same station used for the volumetric capture and reconstruction processes or on a different station with similar characteristics (see Figure 2). The same station has been used throughout this work. Likewise, user can participate in the shared virtual experience by wearing HMDs (e.g., Oculus Rift or Quest) or by using traditional 2D screens.

## 4. Evaluaton

This section first describes our evaluation methodology, setup, and scenario. Then, it presents the results obtained from both objective and subjective tests. In terms of objective data, we report on the consumption of computational and network resources at the client side as well as on end-to-end delays for the involved media streams. In terms of subjective data, we report on perceived interaction quality, togetherness, and immersion levels, as well as on answers given in semi-structured interviews.

4.1. *Methodology, Test Conditions, Tasks, and Procedures*

The main objective of this experiment is to evaluate the presented real-time holoportation technology's appropriateness and readiness for conducting interactive multi-party holomeetings. Objective tests provide insights into the resources required for running multi-party holoconferencing sessions, as well as into the performance of the developed technology (Section IV.C); while subjective tests determine the developed platform's quality of communication and interaction; levels of presence and togetherness; and the effectiveness, naturalness, and comfort when completing tasks.

In addition, the experiment aims to shed some light on how the presented technology is impacted by the number of participants in holomeeting scenarios. To ascertain this, we devised two test conditions, which we conducted on the same groups of four participants in a counterbalanced manner:

- Test Condition A: Sessions with two users.
- Test Condition B: Sessions with four users.

The participants were recruited in groups of four, and we held two parallel sessions with two users for Test Condition A. Although the platform itself can support up to six simultaneous users with point cloud representations, we opted for sessions with a maximum of four participants to simplify our initial assessment of the user's experience, by using a number that can actually boost interaction and has been shown to have a wide applicability [3, 42].

Regarding the tasks to be completed under each test condition, the participants were asked to play a guessing game along the lines of charades, using the following categories: *1. Films; 2. Animals; 3. Sports; 4. Jobs; 5. Celebrities;* and *6. Cities*. The six categories were always played in the listed order, but split across the two counterbalanced test conditions, thus ensuring that our experiments avoided any impact from presentation order effects. A facilitator introduced these categories to the participants prior to the experiment by giving some examples. In iterative rounds, each participant had to choose one option from each category, which the other participants had to guess. During the experiment, the facilitator communicated with the participants via audio connection to indicate the category for each round. The participants were asked to: i) avoid choosing overly intuitive and simple options; ii) use body language and gestures to provide clues about the selected item; iii) complement the non-verbal hints with short audiovisual aids and/or confirmations, if requested; and iv) after a successful guess, have an audiovisual conversation about why the selected option was important to them and sharing related news, stories, or anecdotes. Through preliminary tests, we discovered that the guessing rounds for each category took around 1–2 minutes; so, in addition to the two categories that we introduced in each test condition, we also provided a third one (i.e., Categories 3 and 6) that could be used in case the others were guessed sooner than expected. With this approach, each test lasted between 8 and 12 minutes.

The motivation behind choosing this type of guessing game rather than conducting a less controlled multi-party conferencing session was manyfold. First, it ensures that the shared experiences are not passive because the participants are engaged in an interactive gamified experience with clear tasks for interaction: selecting items for each specific category, giving clues about them, and guessing them. Second, it ensures that all session participants converse and interact around coherent themes and topics, which also helps for the homogenization and aggregation of results from our subjective evaluations. Third, user interactions are not one-sided, as each participant ultimately takes a turn in the protagonist's role of selecting and performing clues for the items to be guessed in each category. Thus, each user must interact with the others, be observed, and be listened to. This also ensures that participants located both at the front and to the sides will be observed (with potentially poorer perceived visual quality for the latter ones, due to the noticeability of lack of full volume). Fourth, using body language and non-verbal gestures to provide the clues for guessing at each item also helps us evaluate the naturalness and quality of the interaction. This is due to the fact that the focus will not be only on interactions via voice and/or facial expressions, but also body language will become relevant (which acquires much higher relevance in this task than in a traditional conferencing session or in any other task focused purely on multiuser communication). Finally, by adopting a charades type of guessing game, we can also determine whether the proposed technology is appropriate and ready for effectively performing multi-user interactive/collaborative (gamified) tasks without any need for extra technological components that enable interactive features (e.g., shared board, video viewing, or some other). All these aspects and features are considered relevant for shedding some light into the formulated research questions (RQ1, RQ2, RQ3) and obtaining meaningful results, while providing an enjoyable experience to participants.

The evaluation protocol and procedures for our user tests are summarized next.

First, participants were recruited according to the following two criteria:

- They had to be older than 18 years old.

- They had to know each other (to ensure fluid and natural social interactions).

  Second, the tests were conducted in the following steps:

- *Step 1 (~10min).* Participants are welcomed and introduced to the test. They are also informed that their participation is totally voluntary, and that they can leave the experiment at any time and for whatever reason.

- *Step 2 (~5min).* Participants fill in a consent form, a demographic and background information form, and a Simulation Sickness Questionnaire (SSQ) [43].

- *Step 3 (~10min).* Participants are brought to the lab room, where they are equipped with the HMD and headphones, assisted by the facilitator(s) if necessary. At this point, they are also introduced to the VR platform, hardware, and environment, namely by entering a virtual welcome room were an experiment facilitator initially converses with them and provides relevant information about the experiment. This allows participants to become familiar with the VR system, hardware, and environment prior to the experiment, and it also contributes to enhancing their comfort.

- *Step 4 (~12min).* When all participants are ready, the facilitators launch the experience. Each session is assigned an id, starting from 1 and increasing the id in 1 for each session. Sessions with odd ids begin with Test Condition A, while even sessions begin with Test Condition B. This counterbalancing method allows avoiding order effects in the obtained results.

- *Step 5 (~15min).* With the help of the facilitator(s), the participants step out of VR and fill in two questionnaires: a version of the social VR experience questionnaire designed in [19], which we slightly adapted by re-phrasing the question items according to the experience being evaluated (see Tables III–VII); and the SSQ [44] questionnaire. This also served as a break from the VR experience.

- *Step 6 (~12min).* Participants undergo the second test condition, as in Step 4.

- *Step 7 (~15min).* Participants fill in the social VR experience and SSQ questionnaires for the second evaluated condition, as in Step 5.

- *Step 8 (~15min).* Participants fill in the System Usability Scale (SUS) [44], NASA Task Load Index (TLX) [45] questionnaire and an additional ad hoc questionnaire for the whole experience (introduced later).

- *Step 9 (~15min).* The facilitators conduct a semi-structured interview in which the participants from each session discuss the presented social VR technology, the experience itself, and other potential applications.

- *Step 10 (~2min).* Participants are thanked, asked for their willingness to participate in future experiments, given a voucher for 30 euros, and allowed to leave.

All questionnaires were administered on paper. Overall, the experiments lasted between 90 and 120 minutes for each session of four participants.

*4.2. Setup and Scenario*

The experiment was conducted in a lab environment with four distributed rooms, whose facilities are shown in Figures 1 and 3. The lab rooms had no background or surrounding noise, appropriate lighting and temperature, and included the necessary equipment for setting up the holomeeting scenarios in groups of four participants. In particular, each room was equipped with a VR-ready laptop with enough computational resources (see Table I) to run the developed technology and scenario, along with a capture sub-system with a single Azure Kinect camera. Despite the limitations of not being able to capture the full volume of a user's body, a single camera setup minimizes the deployment and computational costs, and it can potentially become an appropriate solution in scenarios that do not strictly require 6 degrees of freedom (6DoF), as is the case for this study. As mentioned previously, the point cloud streams were encoded at 15 fps with approximately 50K points per frame (depending on the captured scenes and background removal processes). The client PCs used in the experiment were connected to an Ethernet switch with a capacity of 200 Mbps full duplex. However, the orchestrator used exchange streams between the clients in the same shared session (Table I) was deployed in a data center about 50Km away, thus recreating inter-city social VR sessions via conventional Internet connections, although the participants were placed in rooms within the same building.

During the tests each participant was sitting in a chair and wearing an HMD (Oculus Quest 2 connected to the PC via an Oculus link cable) with an integrated microphone for audio capture, along with noise-cancelling headphones to isolate external noise and have a better perception of spatial audio.

Three experiment facilitators assisted the participants and controlled the test, using chat tools to enable communication between them. Only one experiment facilitator was present in each lab room before and after each test. The orchestrator was used to synchronously launch the shared VR experience for each participant.

The holomeetings were conducted in a newly created 3D meeting room, in which users sat around a round table, as shown in Figure 4 (note the circles on the floor indicating the users' positions). Although a virtual screen is available in the 3D environment, it was switched off (i.e., no content was presented on it) during the experiment. By doing this, the focus is placed exclusively on the end-user representations and audiovisual communication among them, not on any content being consumed together. Once participants were immersed in the 3D environment and the volumetric video and audio communication pipelines had been set up, they began to play the guessing games described in the previous subsection.

Figure 1 provides a screen capture of a holomeeting session with four participants in the 3D virtual meeting room, with two of them represented as point clouds. Figure 5 shows additional captures of the experience, including one of the (self-)views of participants inside the virtual environment.

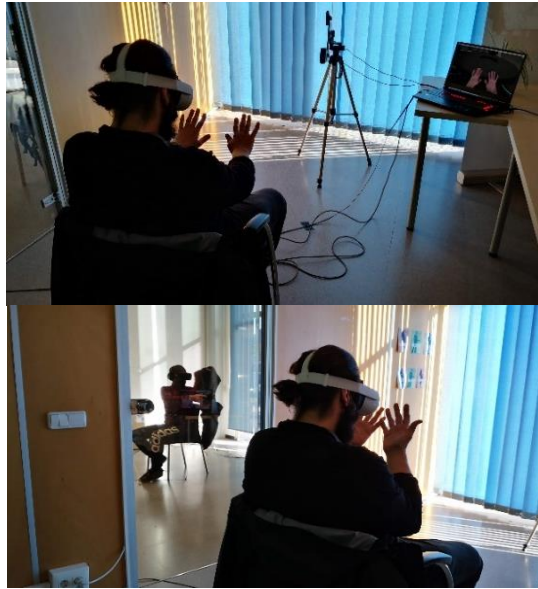

*Figure 3. Setup and distribution of the VR lab facilities where the experiment was conducted*

TABLE I
CHARACTERISTICS & RESOURCES OF THE PCS USED IN THE EXPERIMENT

| PC | CPU | GPU | RAM |
|---|---|---|---|
| **PC_clients** | Intel(R) Core(TM) i7-10750H @ 2.60GHz 2.59 GHz | NVIDIA GeForce RTX 2070 | 16 GB |
| **PC_orchestrator** | Intel Xeon CPU E5-2650 v3 @2.30GHz | NVIDIA Tesla k40m | 64 GB |

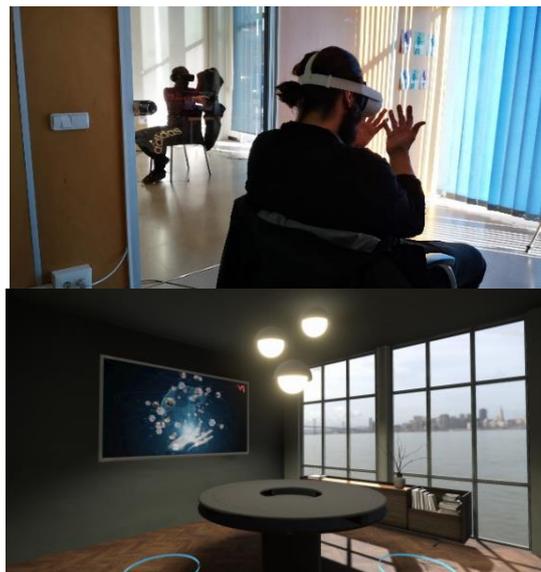

*Figure 4. Designed and produced 3D meeting room for conducting the experiment*

A demo video of the presented social VR platform's capabilities has been made available by the authors: https://www.youtube.com/watch?v=H5VH9Y0FWKE

Likewise, some scenes from the experiments conducted with four users represented as single-cam point clouds can be watched at: https://youtu.be/cvhNsvARNV0

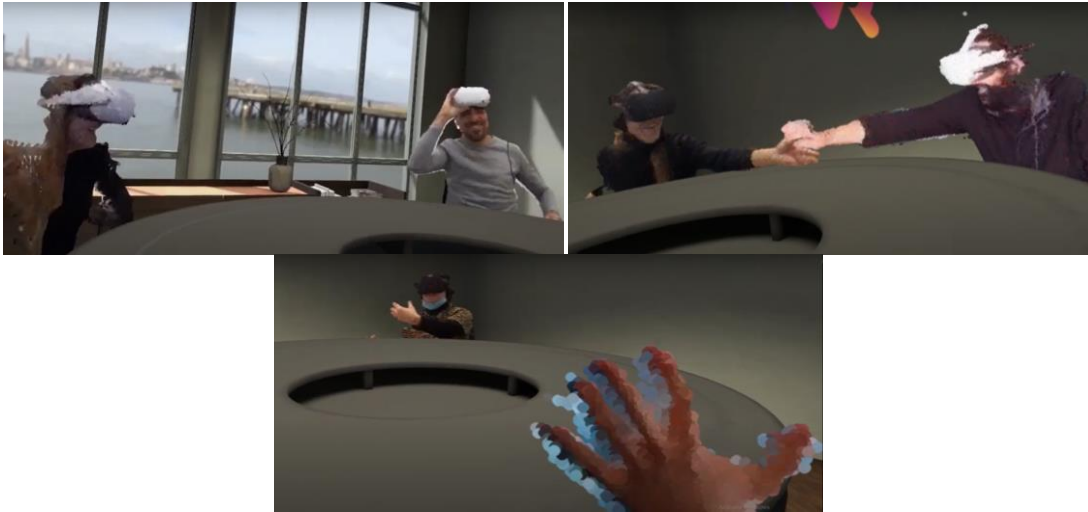

*Figure 5. Screen captures of the experience, showing the participants' appearance and (self-)views inside the virtual meeting room*

4.3. *Results from the Objective Evaluation*

This subsection reports on the objective performance metrics of the client application (i.e., the Unity-based player, presented in Section III.C) when running the experience. In particular, it reports on:

- Computational resources metrics: CPU load (%), GPU load (%), and RAM usage (MB), measured with the tool from [46].
- Bandwidth consumption (Mbps), as reported by Wireshark.[9]
- End-to-end delays, by comparing the capture and rendering timestamps (explained later).

The metrics were measured on the client PCs (Table I) and were sampled throughout the duration of all sessions. The reported values refer to the mean values from five repetitions for each assessed test condition.

Note that the metrics are not reported for the orchestrator, which was run on another PC (Table I) as it only had to manage sessions and forward streams, but not actually process them. Future work will focus on determining its scalability in multi-session setups.

4.3.1. Computational Resources Usage

The usage of computational resources was measured for both test conditions (scenarios with two and with four users) to gain insights into the computational cost of adding more users to the shared experience. Table II summarizes the obtained results, which confirm the initial assumptions of higher consumption of CPU, GPU, and RAM resources when adding extra users. The overall mean usage for the four-participant experience was still far below the upper limits, resulting in a smooth performance that left some margin to add at least one or even two extra point cloud users to client PCs with similar capabilities. However, fluctuations and peaks (Table II) in the resource usage might result in performance issues that could affect the overall experience, especially if adding extra users. The results show that the system is able to balance the resources between CPU and GPU, keeping their usage within safe and quite stable levels, except for some sporadic peaks (e.g., due to intense movements or activity in the VR scenario). Although these results suggest that laptops/PCs with non-dedicated graphics cards could have been used, preliminary tests did show that their computational/graphics resource usage reached levels close to the limits when adding more than one point cloud user representation in a shared session, especially due to sporadic peaks. Therefore, we restricted our study to using VR-ready PCs (Table I) that effectively support scenarios with two and four users.

---
[9] Wireshark, https://www.wireshark.org/ Last accessed in June 2022.

Future studies will include tests on scalability and on heterogeneous client devices to determine the limits, with and without potential optimization solutions.

4.3.2. Bandwidth Consumption

The bandwidth consumption for each point cloud stream (15fps, with around 50K points per frame) from each of the participants was measured using Wireshark. As shown in Figure 6, the TCP throughput for each point cloud stream was quite stable, on the order of 5–7 Mbps (avg=6.2 Mbps, stdv=1.3 Mbps). An analysis of the TCP streams from each point cloud stream in the conducted sessions additionally reflects that TCP errors, re-transmissions, and out-of-order TCP segments were very scarce in the analyzed session (although this depends on the specific network infrastructure and active traffic). This also proves the smoothness and continuity of the media playout for the end-user representations.

These bandwidth consumption values per stream are a bit higher than typical bitrate targets and requirements in high definition (HD) multi-party 2D videoconferencing (around 4 Mbps, according to [47]), but they are comparable to those on video streaming platforms (from 2 Mbps up to 15 Mbps, according to [48]) when delivering HD video (1920×1080) with similar encoding settings. Taking into account that the streams analyzed in this work carry out immersive media content (i.e., volumetric video), these are reasonable and satisfactory bandwidth requirements.

4.3.3. End-to-End delays for point cloud streams

To accurately compute the difference between the rendering and capture instants, end-to-end delays for the point cloud streams were measured by inserting absolute timestamps for each frame captured at the origin side, extracting them prior to rendering the frames at the destination side, and synchronizing the machines by using Network Time Protocol (NTP) at an accuracy of within a few ms [40]. The results show that the average end-to-end (or more accurately capture-to-render) delays for the point cloud streams were:

- Test Condition A (two point cloud streams): 180.5 ms (stdv=11.3 ms);
- Test Condition B (four point cloud streams): 251.2 ms (stdv=12.5 ms).

TABLE II
COMPUTATIONAL RESOURCES USAGE

| Test Condition (TC) | CPU (%) Mean/Max | GPU (%) Mean/Max | RAM (MB) |
|---|---|---|---|
| **TCA**: 2 single-cam point cloud users | 25.5/42.0 | 31.6/62.1 | 843 |
| **TCB**: 4 single-cam point cloud users | 37.8/44.2 | 66.1/69.5 | 846 |

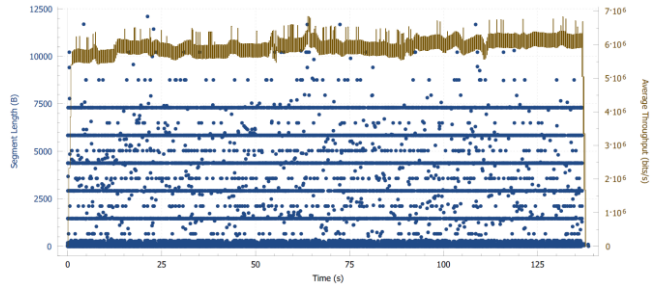

*Figure 6. Average Throughput (bps) and TCP segment length related to a single camera point cloud stream (2 min sample for better appreciation of the temporal fluctuations)*

Therefore, the number of active streams (two and four in Test Conditions A and B, respectively) did have a slight impact on the magnitudes of the delays, mainly due to the higher processing (encoding/decoding) demands once adding extra streams. Likewise, no significant delay differences were noted between the different streams involved in each session, although neither inter-source nor inter-client media synchronization solutions were adopted in this work to compensate for such potential differences [40]. However, none of the test conditions resulted in the magnitudes of the end-to-end delays for each stream exceeding the human tolerance limits for the lack of inter-media and inter-destination synchronization [13, 40]. Therefore, based on the literature [13, 40], these magnitudes of delays do not negatively impact the user experience in interactive multi-party meetings. This eliminates the strict necessity for such media synchronization solutions for conducting interactive meetings. These assumptions will then be confirmed with the results of the user tests.

### 4.4. *Results from the Subjective Evaluation: Sample of Participants*

Overall, 32 users participated in the study, with 17 of them being female and all between 18 and 40 years of age (average of 24.3, standard deviation of 6.7). Twenty-five of them (78.1%) were right-handed, six left-handed, and one was ambidextrous. None of them expressed having audio-visual impairments. These data were extracted from the demographic questionnaire.

From the background questionnaire, four participants (12.5%) declared having a novice level of computer skills, twenty (62.5%) intermediate, and eight advanced. The questionnaire asked whether they had any previous experience with VR and social VR tools. Half of them stated having had previous VR experiences, while only four (12.5%) declared having had previous experiences with avatar-based social VR platforms. To overcome this lack of experience, all participants were introduced to the VR platform, hardware, and environment prior the experiment, as detailed in Section IV.A.

### 4.5. *Results from the Subjective Evaluation: Answers from Questionnaires*

4.5.1. Results from the Social VR Experience Questionnaire

After each test condition, the participants were asked to complete the social VR experience questionnaire. As detailed in [19] and [5], the questionnaire includes items about emotions, feelings, perceptions, and opinions regarding crucial aspects of social VR experiences, which we categorized into four main sections (whose results are in Table III-XI):

- Quality of interaction (QoI): the emotional experience, quality of communication, and naturalness of communication.
- Social meaning (SM) or connectedness: the feeling of togetherness, of emotional closeness, and the enjoyment of the relationship.
- Presence / Immersion (PI): the plausibility and illusion of space.
- Additional aspects: realism, how much the users like the content, spatiality, etc.

Most answers for each part of the questionnaire were on a 5-point Likert scale. The possible answers are detailed in Table III. The acronyms in that table are also used in the next tables (Tables IV–VII), which provide the results from each part of the social VR experience questionnaire.

The extra questions (Table VII) from the social VR experience questionnaire were asked only once, after the participants finished both test conditions. Tables IV–VI and VIII report only the results for the sessions with four participants, because: 1) having extra participants is more technologically challenging; and, especially, 2) the results obtained for both test conditions are very similar, as Wilcoxon–Mann–Whitney tests (procedure explained in [5]) found no significant differences in the QoI, SM, and PI parts of the questionnaire between the two test conditions.

The obtained results are highly satisfactory for each part of the social VR experience questionnaire. This reveals that the presented technology and platform can provide high levels of QoI, SM, and PI. The participants were also very satisfied with the quality, realism, and spatiality of the created VR scenario.

In addition to the questions listed in Table V, Q23 asked *how emotionally close each user felt to the other users*, using a 7-point scale with two circles separated by different distances (from separated to totally overlapped), as seen in Table VIII. The results found that the majority of answers given were options 5 and 6, showing significant overlapping of the circles and thus allowing us to conclude that participants did indeed feel quite emotionally close, which also indicates feeling togetherness and intimacy.

4.5.2. Results from the SSQ Questionnaire

With regard to the SSQ, none of the respondents reported any of the test conditions causing significant effects or symptoms as a result of the experiences. This was also confirmed during the semi-structured interviews that asked about this explicitly. In addition, none of the participants reported having felt dizziness or becoming tired as a result of the two test conditions.

4.5.3. Results from the SUS Questionnaire

With regard to usability, the average SUS score obtained was 91 (Figure 7). This score is significantly above the average of 68 in other studies [44], and it corresponds to a letter grade of A+ (top score grade).

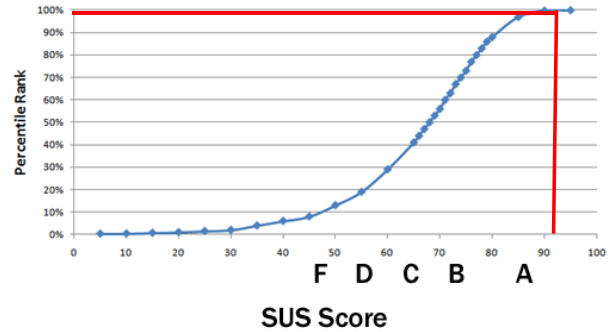

Figure 7. SUS score and percentile rank for the social VR platform used to conduct the holomeeting experiments

TABLE III
5-POINT LIKERT SCALE FOR RATING THE ITEMS OF THE SOCIAL VR QUESTIONNAIRE

| Acronym | Meaning | Assigned Score |
|---|---|---|
| TD | Totally Disagree | 1 |
| PD | Partially Disagree | 2 |
| NN | Neither Agree nor Disagree | 3 |
| PA | Partially Agree | 4 |
| TA | Totally Agree | 5 |

TABLE IV
RESULTS FROM THE SOCIAL VR EXPERIENCE QUESTIONNAIRE (HOLOMEETINGS EXPERIMENT) – QUALITY OF INTERACTION (QoI)

| Questions | TD | PD | NN | PA | TA |
|---|---|---|---|---|---|
| Q2. I was able to feel the other users' emotions in the shared VR scenario. | - | - | 2 | **16** | 14 |
| Q3. I was sure that the other users often felt my emotions. | - | - | 4 | **18** | 10 |
| Q4. The virtual experience with the other users seemed natural. | - | - | - | **23** | 9 |
| Q5. The actions used to interact with the other users were similar to those in the real world. | - | - | 2 | **19** | 11 |
| Q6. It was easy for me to contribute to the conversation. | - | - | 1 | 8 | **23** |
| Q7. The conversation with the other users seemed highly interactive. | - | - | 1 | **16** | 15 |
| Q8. I could readily tell when the other users were listening to me. | - | - | 1 | 12 | **19** |
| Q9. I found it difficult to keep track of the conversation. | **19** | 11 | 2 | - | - |
| Q10. I felt completely absorbed in the conversation. | - | - | - | 12 | **20** |
| Q11. I could fully understand what the other users were talking about. | - | - | - | 11 | **21** |
| Q12. I was very sure that the other users understood what I was talking about. | - | - | 1 | 13 | **19** |
| Q13. I often felt as if I was all alone in the virtual experience. | **26** | 6 | - | - | - |
| Q14. I think the other users often felt alone in the virtual experience. | **22** | 10 | - | - | - |

TABLE V
RESULTS FROM THE SOCIAL VR EXPERIENCE QUESTIONNAIRE (HOLOMEETINGS EXPERIMENT) – SOCIAL MEANING (SM)

| Questions | TD | PD | NN | PA | TA |
|---|---|---|---|---|---|
| Q15. I often felt that the other users and I were sitting together in the same space. | - | - | - | 12 | **20** |
| Q16. I paid close attention to the other users. | - | - | - | **18** | 14 |
| Q17. The other users were easily distracted when other things were going on around us. | 9 | 7 | **14** | 1 | - |
| Q18. I felt that having the VR experience together enhanced our closeness. | - | - | 3 | **21** | 7 |
| Q19. Having the VR experience together created a good shared memory between us. | - | - | - | 16 | 16 |
| Q20. I derived little satisfaction from the virtual shared experience. | **16** | 12 | 4 | - | - |
| Q21. The VR shared experience with my partner felt superficial. | 10 | **15** | 6 | - | 1 |
| Q22. I really enjoyed the time spent with the other users. | - | - | - | 6 | **22** |
| Q24. In the virtual world I had a sense of "being there". | - | - | - | 12 | **20** |
| Q25. Somehow, I felt that the virtual world was surrounding me and my partner. | - | 1 | 3 | 12 | **16** |
| Q26. I had a sense of acting in the virtual space rather than operating something from outside. | - | - | **3** | 11 | **16** |
| Q27. My virtual shared experience seemed consistent with a real-world experience. | - | - | 6 | **18** | 8 |
| Q28. I did not notice what was happening around me in the real world. | - | 3 | 2 | **15** | 11 |

TABLE VI
RESULTS FROM THE SOCIAL VR EXPERIENCE QUESTIONNAIRE (HOLOMEETINGS EXPERIMENT) – PRESENCE/IMMERSION

| Questions | TD | PD | NN | PA | TA |
|---|---|---|---|---|---|
| Q29. I felt detached from the outside world while having the VR experience. | - | - | 9 | 9 | **14** |
| Q30. At the time, the shared VR experience with the other users was my only concern. | - | - | 5 | 13 | **14** |
| Q31. Everyday thoughts and concerns were still very much on my mind. | **11** | 9 | **11** | - | 1 |
| Q32. It felt like the VR shared experience took less time than it actually did. | - | - | 3 | 12 | **17** |
| Q33. When having the VR experience together, time appeared to go by very slowly. | **20** | 8 | 4 | - | - |

TABLE VII
RESULTS FROM THE SOCIAL VR EXPERIENCE QUESTIONNAIRE (HOLOMEETINGS EXPERIMENT) – EXTRA QUESTIONS

| Questions | TD | PD | NN | PA | TA |
|---|---|---|---|---|---|
| Q34. I liked the created VR scenario for holding virtual meetings. | - | - | 1 | 14 | **17** |
| Q35. The created VR scenario is realistic (i.e., resembles a real scenario). | - | - | 4 | **15** | 13 |
| Q36. The spatiality in the VR scenario (i.e., perceived distances and sizes of elements, including the participants' bodies) is consistent with a real-life scenario. | - | - | 5 | 11 | **16** |

TABLE VIII
EMOTIONAL CLOSENESS BETWEEN PARTICIPANTS (Q23)

| 1 | 2 | 3 | 4 | 5 | 6 | 7 |
|---|---|---|---|---|---|---|
| 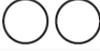 | 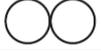 | 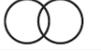 | 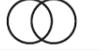 | 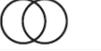 | 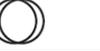 | 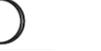 |
| - | - | 6 | 2 | 10 | 9 | 5 |

TABLE IX
RESULTS FROM THE AD HOC QUESTIONNAIRE (HOLOMEETING EXPERIMENT) – PART 1: AUDIOVISUAL QUALITY

| Questions (Rating Scales: 1 = bad, 2 = poor, 3 = fair, 4 = good, 5 = excellent) | 1 | 2 | 3 | 4 | 5 |
|---|---|---|---|---|---|
| Q37. The visual quality of the virtual scenario | - | - | 1 | **22** | 9 |
| Q38. The visual quality of my representation | - | 2 | 13 | **15** | 2 |
| Q39. The visual quality of the representation of the user(s) next to me | - | 4 | **16** | 12 | 0 |
| Q40. The visual quality of the representation of the user(s) in front of to me | - | - | 7 | **21** | 4 |
| Q41. The audio quality from the user(s) next to me | - | - | - | **19** | 13 |
| Q42. The audio quality from the user(s) in front of to me | - | - | - | **20** | 12 |

TABLE X
RESULTS FROM THE AD HOC QUESTIONNAIRE (HOLOMEETING EXPERIMENT) – PART 2: COMPARISON TO A REAL SCENARIO

| Questions (Rating Scales: 1 = much worse, 2 = slightly worse, 3 = equivalent, 4 = slightly better, 5 = much better) | 1 | 2 | 3 | 4 | 5 |
|---|---|---|---|---|---|
| Q43. The visual quality of the virtual scenario compared to a real one | - | 12 | **16** | 4 | - |
| Q44. The overall experience with two users compared to the one with four users | - | 2 | **15** | 13 | 2 |
| Q45. The overall virtual experience compared to one in real life | - | 2 | **19** | 3 | 1 |
| Q46. The visual representation of users in the virtual experience compared to a real scenario | 2 | **19** | 11 | - | - |
| Q47. The audio quality in the virtual experience compared to a real scenario | - | 5 | **20** | 7 | - |
| Q48. The naturalness of the gestures in the virtual scenario compared to a real scenario | - | 8 | **22** | 2 | - |
| Q49. The fluidity of the gestures in the virtual scenario compared to a real scenario | - | 11 | **19** | 2 | - |
| Q50. The overall communication quality in the virtual scenario compared to a real scenario | - | 5 | **19** | 7 | 1 |

TABLE XI
RESULTS FROM THE AD HOC QUESTIONNAIRE (HOLOMEETING EXPERIMENT) – PART 3: POTENTIAL AND IMPACT

| Questions (Rating Scales: 5 = Strongly Agree, 4 = Agree, 3 = Neutral, 2 = Disagree, 1 = Strongly disagree) | 1 | 2 | 3 | 4 | 5 |
|---|---|---|---|---|---|
| Q50. This system is effective for holding virtual meetings. | - | - | - | 6 | **26** |
| Q51. The distances between the participants were appropriate. | - | - | 5 | 8 | **19** |
| Q52. The quality of the user representations is enough to enable effective and comprehensive interactions and collaborations. | - | 1 | 1 | **21** | 9 |
| Q53. Wearing an HMD prevented an effective and satisfactory interaction experience (due e.g., inability to see each other's eyes). | 6 | 6 | **16** | 4 | - |
| Q54. I would use a system like this for meetings and collaborative tasks in virtual scenarios with up to four users. | - | - | - | 10 | **22** |
| Q55. I would use a system like this for meetings and collaborative tasks in virtual scenarios with more than four users. | - | - | 4 | 13 | **15** |
| Q56. These kinds of systems can contribute to a more sustainable environment. | - | - | 5 | 12 | **15** |
| Q57. These kinds of systems can contribute to saving time and money. | - | - | 1 | 8 | **23** |

4.5.3. Results from the Ad-hoc Questionnaire

An ad hoc questionnaire was also designed to capture insights into additional features and aspects of the evaluated technology, setup, and scenario after having experienced both test conditions. The different aspects to be evaluated through the ad hoc questionnaire and results obtained for each of the associated question items are provided in Tables IX–XI. The ad hoc questionnaire was specifically designed to measure users' perceptions and opinions with regard to: i) the audiovisual quality of the user's representations; ii) how the recreated virtual meetings and scenarios compare to real ones; and iii) the potential impact of the presented technology and use case.

The first part of the ad hoc questionnaire (Table IX) focused on the perceived audio-visual quality. In general, participants were quite satisfied with the visual quality of the scenario and with the audio quality from all users, regardless of their positions. When referring to the visual quality of the end-user representations, participants were quite happy with theirs and those of the others, especially noting a better quality for the users located in front of them. This is reasonable, as the experiment used a capture system with a single camera placed in front of users. Therefore, the full volume was not captured, especially with regard to the lateral and posterior parts of the body. Still, the quality was considered acceptable by the participants in such cases.

The second part of the questionnaire (Table X) focused on comparing the perception of the virtual scenario/experience to a real one. The visual quality of the VR scenario, as well as the audio quality and naturalness of the gestures in the virtual scenario, were generally rated as equivalent to those in a real scenario. The responses were nearly the same regarding the fluidity of gestures, although many users rated this as slightly worse than in real scenarios. Most of the participants rated the overall communication quality and virtual experience as equivalent to those in real scenarios. Likewise, and interestingly, most of the participants rated the overall experience with two users as equivalent (46.9%) or slightly better (40.6%) to the one with four users. On the one hand, this somewhat confirms that the perceived performance of the system with four users was still satisfactory. On the other hand, it may also be proof that having more users also increases the interaction possibilities, despite the noticeable limitations perceived in the visual quality of users sitting to one's side (i.e., side viewpoints) in experiences with four users.

The third part of the questionnaire (Table XI) assesses the potential and impact of the developed technology and evaluated scenario. Most of the participants believed the system is viable for holding online virtual meetings. What is more, the participants generally agreed that the distance between them in the virtual scenario was appropriate. Despite the noticeable visual quality limitations of the user representations, the participants generally believed that the current quality is sufficient for effective and comprehensive interactions and collaborations. Wearing an HMD was still perceived as inconvenient for some participants, but not as a limiting barrier (in the interviews conducted at the end of the experiment, most participants affirmed having felt comfortable during the experience. In general, participants seem to be very interested in using systems like the one evaluated for holding online meetings, including scenarios with up to four and even more than four users. Finally, most participants believed that these kinds of systems can contribute to a more sustainable environment, as well as to saving time and money.

*4.6. Results from the Subjective Evaluation: Answers from Interviews*

After having finished experiencing the two test conditions and filling in the associated questionnaires, the four participants from each session took part in a semi-structured interview with the experiment facilitators. The audio recordings of the interviews were transcribed and coded, following an open coding approach [49]. Since the interviews were conducted with all the participants from each group together, their answers have been transcribed and coded both as individual participants (labelled P1–P32) and as groups (labelled G1–G8). The answers and items included in the interview can be derived from the answers reported in the next subsections, which have been structured based on the key themes and aspects that emerged.

4.6.1. Key Impressions

Participants were first asked their impressions, and keywords were compiled for how they described their virtual meeting experience. The following keywords can be highlighted: impressive (46.9%), futuristic (43.8%), innovative (43.8%), amazing (31.3%), funny (31.3%), wow! (25%), next-generation video conferencing (25%), interesting (25%), surprising (21.9%), technology that enables new possibilities (15.6%), and incredible (12.5%).

4.6.2. Benefits and Potential of Social VR

All participants thought that the presented social VR platform allowed them to experience "social presence", having felt identified with the end-user representations, including theirs and the other's representations. "*It was not an avatar, but me!*", as stated by P4, P9, P19, and P28 (three of whom reported having previous experiences on social VR platforms). The participants generally reported "feeling together" with the others, which enriched the overall experience. G1, G4, and G7 stated, "*We felt together, sharing an activity and experience, and this is really an added value to VR.*" G2 expressed, "*That is super! You can meet with whoever you want, anywhere, anytime!*" And G6 mentioned, "*This really gives the feeling of being in the same place, together.*" Both the scenario and experience were found to be immersive by all participants.

The participants in general felt comfortable in the virtual environment. A few participants (12.5%) – interestingly, all of them without previous VR experience – mentioned feeling a bit tense at first contact with the social VR platform, because of the uncertainty, but they then quickly felt more relaxed. None of the participants felt uncomfortable wearing the HMD, and many of them (37.5%) even stated they forgot they were wearing it during the experiment.

The quality of communication and interaction was also found satisfactory in general. Even though visual artefacts were noticed, the participants found it impressive to see themselves and the other participants integrated into the shared virtual environment. "*I could even see my tattoos and the details of my clothing,*" participants from G3 and G5 stated. Participants also pointed out that the perceived delays were a minor issue (25%) and that although facial expressions were partially blocked by the visual quality and the HMD occlusion (31.3%), all of them stated that having realistic and volumetric representations of users is impressive and enabled natural and rich interaction. None of the participants reported any lack of fluidity or audio artefacts during the experiences. "*It was clear that the audio was directional, as you could identify from where it came*", stated users from G2 and G8. In general, the interactions between the participants were perceived as natural. "*The interaction was natural. It is not the same as in a real scenario, as eye contact is missing, but we were able to effectively communicate and perform the requested tasks clearly and effortlessly*", as stated by G1 and G5. Indeed, none of the participants reported any problems with communication and interaction during the test sessions.

Participants also found limitations related to having used a single-camera capture system in the experiment. First, the representation quality of users placed next to them was generally perceived to be lower than for users in front of them. Second, artefacts in the self-representations were also mentioned by G1, 63, and G7, especially with regard to the arms, hands, and feet. P7, P11, and P26 mentioned: "*The quality of my partner's representation seemed better than mine*". Three participants (P1, P13, P31) suggested decreasing the point sizes in the end-user representations (which is a technical setting of the point cloud representation format feasible to adjust).

All participants believed that the volumetric representations for the end users can help maintain, strengthen, and even create new relationships in real life. G1, G3, and G6 (including participants with and without previous VR experience) stated, "*It is a very innovative and useful solution. We have friends and family members living apart. This would enable us to meet and share experiences, overcome distance barriers, and save time*". In general, participants believed that these systems can be applied to interact with both known people and new contacts. The suggested applicability use cases for this social VR technology are enumerated later.

Half the participants (50%) affirmed it was an amazing experience for them, and over a third (37.5%) suggested that social VR can be a powerful tool for avoiding the real world in certain situations. Seven of the eight groups remarked that it could be an especially valuable communication medium for overcoming the social distancing measures brought on by the pandemic.

4.6.3. Weaknesses and Missing Aspects in Social VR

Many participants (62.5%) explicitly mentioned that higher visual quality would be more desirable, especially for the self-representation and the representations of the lateral users. "*The current level of quality is enough to effectively communicate and interact, but this is not yet like real-life scenarios*", stated four of the groups. "*The quality of the end-user representations should improve in the future*", declared G7. None of the participants reported lack of fluidity or any delays affecting the overall experience.

Integration of multi-sensory stimuli like scents (12.5%) and especially haptic feedback (75%) was identified as a missing aspect. G1, G3, and G8 stated: "*It would be great if you could touch things, and if the haptic interactions indeed had an effect on the VR environment or story*". "*Having the table gave the impression that I could pick up objects from it, or leave objects on it*", stated three participants. Participants in four of the eight sessions tried to shake hands in the virtual scenario.

75% of participants would also like to move freely in VR (e.g., 6DoF). "*It would be great if you *could move around, get closer to elements and participants in the shared environment*", stated participants in G2 and G4.

Participants envisioned more interactive and active experiences thanks to a future combined support of haptic feedback and 6DoF features: "*If you can actively explore things and complete collaborative tasks together, as well as influence the VR environment, then you would be able to perform very useful activities through social VR*", remarked P2 and P25.

4.6.4. Weaknesses and Missing Aspects in Social VR

In general, the participants foresee a big impact from social VR in the near future. They identified the following use cases as the most interesting for social VR: meetings (87.5%), virtual conferences (60%), gaming (50%), training and education (37.5%), virtual consultation (37.5%), dating (25%), and shared video watching (25%). No clear correlations have been detected between the user's profiles and backgrounds for these opinions.

Participants believed that social VR is a powerful medium for meeting not only known users, but also new contacts. All participants showed interest in using social VR in the future. "*It would be great to have such a system at home!*" stated members of many of the groups. "*This is the next generation meeting tool*", stated P5 and P16. Participants in two of the groups were also concerned about the price of the technology: "*Of course, this technology is great, but adoption will depend on its cost*". A few participants (12.5%) also declared that social VR might be more adequate in corporate environments in the short term, and not yet for domestic environments. Others (12.5%) showed concerns about social VR contributing to sedentariness.

All participants believed that having more than two participants in the same session is very useful and provides added value, although the visual representation of users placed at the sides would need to improve so the full body volume can be seen. This is in line with the results from the ad hoc questionnaire (Tables IX–XI).

4.6.5. The Next Generation of Social VR

The next generation of social VR is envisioned by participants as:

- Higher quality volumetric representations, ideally not blocking the facial expressions (i.e., by removing the HMD) (75%).
- Scenarios with 6DoF and no cables (62.5%).
- eXtended Reality (XR) environments that blur the boundaries between the real and virtual worlds (50%).
- Multi-sensory environments that stimulate all the senses, especially touch and, ideally, smell (37.5%).

Two groups also suggested adding a feature to personalize the meeting environment and/or even importing the desired one. "*It would be great if you could choose the place where you want to meet, and personalize it*", stated P8 and P29.

## 6. Discussion

This work has shown the applicability, viability and readiness of a novel lightweight and low-cost holoportation technology for effectively holding interactive multi-party holoconferencing services (**RQ1**). In doing so, we have assessed how the number of users impacts the performance and user experience, specifically by conducting sessions with two and four users (**RQ2**). Next, we discuss the implications of the obtained results with regard to these two research questions.

In terms of performance, we have shown that the presented platform is able to hold smooth and fluid multi-party sessions with up to four distributed users, as well as with up to six users in controlled lab scenarios under stable conditions. In the latter case, unstable performance could be provoked by computational peaks and overloads due to other running applications or intense activity (e.g., highly dynamic movements) in the holomeeting scenarios, together with varying network conditions, which could result in noticeable audiovisual artifacts or even service outage. For this reason, we chose to study groups of up to four users, which is indeed supported by previous studies showing that this number of users boosts high interaction levels and is widely applicable [3, 42]. In the experiments with two- and four-participant sessions, the peaks in computational resource usage never reached levels that compromised performance and smoothness (e.g., large delays, image and sound artifacts, decrease in fps, etc.). In addition, the results show that end-to-end delays slightly increased when increasing the participants from two to four, as expected, but the delays were still on the order of 200ms, which can be considered very satisfactory, as these values are significantly below the maximum thresholds for effective interactive communications [40, 49]. Finally, we have shown that each point cloud stream (with approximately 50K points per frame and 15 fps) required around 5–7 Mbps, with no major fluctuations in the data rates and no occurrence of TCP errors in the conducted experiments. These bandwidth requirements are slightly above those in HD conferencing systems [47, 48], but they are reasonable, given the fact that immersive media content is being exchanged. For large-scale conferencing sessions, scalability optimizations would need to be devised, but that is also the case for processing requirements.

In terms of user experience, the participants in general reported being highly satisfied with regard to usability, (co-)presence, interaction quality, workload, and task effectiveness, which are essential aspects in multi-user (VR) scenarios (e.g., [5], [9], [19]). The participants also expressed being highly satisfied, impressed, and interested in the presented technology and the scenarios it enables. Although half of the participants indicated having had no previous experience with social VR tools, an average SUS score of 91 was obtained. This score is clearly above average and indicates that that the system fits well to the purpose for which it was built. Furthermore, none of the participants declared feeling dizzy or becoming tired as a result of experiencing the two test conditions, and none of them reported having any troubles or difficulties in conducting the tasks at hand. To the contrary, they reported being able to perform these tasks effortlessly and comfortably.

Finally, the results from the social VR questionnaire are very positive for QoI, SM, and PI. The analysis of the obtained results per test condition provides relevant insights into both **RQ2** and **RQ3**. The fact that no significant differences were found to exist between the two-participant and four-participant test conditions suggests that both scenarios were well received and awakened interest, which is in line with previous research works [3, 42]. Despite this, further relevant insights have been gathered from specific ad hoc question items regarding these research questions. On the one hand, nearly half (46.8%) of the participants stated that the overall experience in the sessions with two users was equivalent to the sessions with four users, and a significant percentage of them (40.6%) declared the two-user sessions to be slightly better. These slightly better ratings for the sessions with two users may be due to the higher perceived quality of visual representations for the users situated to the front (users in two-participant sessions face each other) than for those located to the sides (Table IX). It may also be ascribed to the better performance provided when a lower number of users participate in the shared session. On the other hand, despite these performance issues and especially the implications of using single-camera capture systems for side views (which were confirmed as notable in the results from Table IX and in the interviews), they seemed to have no negative effect when participants were asked about the interaction quality, social meaning, and presence levels, thus demonstrating that these simple and cheap capture systems can be considered valid for scenarios in which participants are not generally expected to navigate within the virtual environment (i.e., 3DoF+). This indicates that the single-camera capture system is acceptable for small groups of users whose relative positions ensure a certain level of frontal visibility, although this may be less appropriate for other environments (e.g., higher number of users, different and varying positions of users in the scene, 6DoF environments) where full volumetric body reconstructions may be required (to be explored in future studies). In relation to **RQ2**, all the participants stated in the interviews that added value and greater utility could be gained from having more than two participants in the same session. This suggests that the slight preferences toward two-user sessions might be due to the higher quality of the frontal users relative to those at the side, and that it is not necessarily due to the number of users per se (**RQ2**).

Overall, the recreated virtual meetings were perceived as equivalent to or at most just slightly worse than an associated real scenario, mainly in terms of perceived audiovisual quality, the naturalness and fluidity of gestures, and levels of realism (Table X). Although a comparison with a baseline condition (i.e., face-to-face meeting) has not been conducted to confirm this, these are very promising results that also inform our research questions.

In a nutshell, both the presented technology and use case have aroused great interest among the participants, thus forecasting relevant benefits and impacts in the near future. Both performance and user experience have been highly satisfactory among the groups of two and four participants, despite the technological limitations from lacking very high resolution and full volumetric representations. These are promising results that reinforce the potential for multi-party holoportation. In pre pre-pandemic times, social VR was perceived as a technology and medium mostly relevant for entertainment and exhibition use cases, as reflected in [5, 7]. However, the arrival of COVID-19 jointly with the evolution of the platform (higher visual resolution and user numbers compared to the solution in [5]) have generated stronger desires and preferences for using this technology in domestic scenarios to meet with others in both personal and professional contexts.

In addition, authors are aware that other types of interactive, collaborative and/or gamified tasks could have been chosen and adopted for conducting the user study. However, as discussed in Section IV.A, the selection of a guessing game along the lines of charades, with up to four users equally distributed around a round table, was intended to accomplish key conditions, features and aspects of the planned evaluations toward obtaining valuable insights into the formulated research questions. Some of these factors include: avoiding one-sided conversations, by adopting a turn-taking basis; ensuring all users are observed and listened to, regardless of their position in the virtual environment; implicitly requiring the use of body language and non-verbal gestures, thus not just relying on interaction via voice and facial expressions; and focus the attention on the user representations and their gestures rather than on the virtual environment or any other content / element presented on it. Even though the task to be conducted is not the main focus of our study, and alternative ones could also serve the purpose (discussed next), authors believe that the adoption of such a multiuser guessing game has been an appropriate decision, having helped to implicitly and naturally gather coherent and valuable results and insights aligned with the pursued objectives. This is corroborated by the results from the Social VR experience an ad-hoc questionnaires, as well as from the interviews, from which the next aspects are highlighted: emotions were felt (Q2, Q3); conversations and interaction actions (including non-verbal gestures) were perceived smooth and natural (Q4-Q5, Q48-Q49), as well as effective and meaningful (Q6-Q12); users felt immersed in the virtual environment (Q24-Q33); users paid close attention to the other users and their actions (Q16-Q17); and users felt together (Q13-Q15, Q23). The adoption of this task has also contributed to gathering coherent and satisfactory ratings for the perceived visual quality of the end-user's representation and the audio from them, from both frontal and side viewpoints (Q37-Q42, Q46-Q47), being such quality levels considered sufficient for holding effective and comprehensive holomeetings (Q50-Q55). In general, users felt comfortable during the experience and enjoyed it (Q18-Q23), even stating that it resembled a physical one (Q24-Q27) and that it seemed to take shorter than it actually did (Q32-Q33). Finally, the fact of having adopted a turn-taking interactive task for conducting the experiment rather than a less controlled conferencing session may have also contributed to gathering valuable inputs regarding suggested extra features and potential applicability scenarios for this technology. Therefore, although the study focuses neither on the task to be conducted nor on comparing the suitability of such a task in comparison with other alternative ones, authors firmly believed that its selection has resulted satisfactory for the evaluation of the presented multi-party holoportation technology, in terms of technological and user experience aspects.

Notwithstanding the highly satisfactory and promising results that have been obtained, some limitations from both the presented technology and study have been also identified.

When it comes to the technological aspects, the main limitations worth mentioning and considering for future studies are:

- Higher visual quality and full volumetric representations: Despite the positive scores for presence, co-presence, and interaction quality, as well as the positive responses from the ad hoc question items regarding perceived visual quality, the visual representation of users still has margins for improvement. Users are generally impressed by the innovative aspect of seeing representations of one's self and of others in real-time and in a shared virtual environment, which may have driven the high scores. However, they also recognized the need for improving the visual quality and for providing full volumetric representations. As mentioned, the current level of detail (LoD) per user representation is 50K points at 15fps. This LoD should be enough to render and display good image quality, but three potential improvements can lead to a significant increase in quality: i) adopting higher quality sensors (with higher resolution and more accurate depth estimation); ii) designing and adopting more powerful and efficient point cloud encoding strategies for real-time multi-party communications; and iii) designing and adopting intelligent fusion and reconstruction strategies for composing the full volume while also cleaning and aligning the captured data. All these optimizations are expected to result in higher visual quality, and thus lead to a better QoE while also overcoming the limitations from the perceived lack of full volumetric representation for side users, as reported in this work.

- Blocked facial expressions and discomfort from using an HMD: As mentioned by participants, even though the current generation of HMDs do not have serious comfort issues, they are still somewhat cumbersome and block eye contact and facial expressions. This may not be a serious drawback in entertainment and collaboration scenarios, where the focus is mainly on the content being watched or on the task to be performed, respectively; but it can be a relevant factor in potentially critical or sensitive person-to-person conversations. The authors are fully aware of this issue, but the scientific community is actively investigating effective solutions for replacing in real-time the HMD with estimated facial expressions (e.g., [50]). In addition, expectations are that the next-generation eXtended Reality (XR) headsets will be more lightweight and, further, they will block facial expressions and eye contact to a much lesser extent.

- Scalability: The scalability of the current system is limited. Although sessions with up to six users have been achieved in stable conditions and controlled environments, adding extra users impacts performance and thus the perceived QoS and QoE. Likewise, it is expected that higher resolution RGB-D cameras will be available in the near future, thus leading to an emerging need for real-time processing of even more points or voxels for each point cloud stream. Finally, it will be necessary to devise and deploy efficient network-based processing solutions like multipoint control units (MCU) [51], remote rendering engines, and smart LoD adjustment solutions. Such developments will need to efficiently adapt to varying conditions and heterogeneous environments, including the usage of client devices with low processing capabilities. Still, given the fact that we have enabled real-time interactive sessions using realistic and volumetric representations for up to four and even six distributed users, this already represents a highly relevant milestone and provides much added value.

When it comes to the study itself and our methodology, the main limitations worth mentioning and considering for future studies are:

- <u>Comparison with a baseline condition and/or alternative test conditions</u>: Although the evaluation material (questionnaires and interviews) explicitly included question items for rating the virtual experience in comparison to a real-world experience, we have not considered any baseline test condition (i.e., physical meeting, use of a traditional 2D video conferencing tool, or of even another social VR platforms using avatars or 2D representations), although other works have done so (e.g., [5], [9] [19]). While the authors are aware that including these conditions would have provided more conclusive insights into the potential benefits and impacts of the presented technology for conducting interactive multi-party meetings, the presented study represents a first – but relevant and necessary – step in assessing and confirming the potential impacts of this novel technology and medium.

- <u>Objective evaluation of point clouds</u>: The conducted experiments have included both objective performance tests (mainly in terms of delays and bandwidth usage measurement) and subjective perception tests for the point clouds (using questionnaires and interviews). In this context, momentum is gaining for the development and adoption of objective quality assessment metrics for point clouds (e.g., [52-54]), but our study has not considered these aspects. Such metrics can provide relevant insights, such as from determining visual quality when adopting different encoding strategies, scenarios, and setups; and they can allow avoiding lack of unanimity among responses while minimizing human intervention. Nevertheless, our study has been conducted in a controlled scenario while adopting the maximum resolution and frame rate granularity supported by the adopted sensors and encoders/decoders [6, 36]; and our evaluations show that the adopted methodology is acceptable and comprehensive enough to obtain valuable insights into the formulated research questions. Still, the objective metrics mentioned above should be incorporated into future studies when considering how to conduct evaluations under different test conditions and repetitive/automated tests.

<u>Investigating other scenarios, metrics, tasks and test conditions</u>: The experiments included test conditions with two and four participants in a specific setting with participants equally distributed around a round meeting table, playing guessing games along the lines of charades. These have been proved to be adequate experiment design decisions, providing relevant insights into the formulated

RQs. However, additional test conditions, tasks and scenarios would need to be evaluated in further studies to confirm the presented technology's viability for effectively holding interactive multi-party holomeetings and collaborative tasks in different contexts and use cases. Likewise, although the experience was found to be highly interactive, with all users having contributed effortlessly and naturally to the selected tasks (according to the results from the subjective evaluation), no metrics on the time spent looking at others and/or talking to others [5], or on tasks' completion times and success rates, have been considered. These types of metrics can become useful to assess and validate the formulated research questions and future ones, and their adoption will be considered in future works.

## 7. Conclusions and Future Work

This article has presented a novel lightweight and low-cost social VR platform that enables interactive multi-party meetings and conferencing sessions with real-time holographic and volumetric user representations, using low-cost and off-the-shelf equipment like VR-ready laptops, Oculus Quest 2 HMDs, and Azure Kinect cameras for deploying the single-sensor capture systems.

Our evaluation of the system performance and user experience in sessions with groups of two and four participants in interactive holomeeting scenarios has provided highly satisfactory and promising results. To our knowledge, this has been a pioneering work that demonstrates the readiness and applicability of holoportation technology for effectively holding interactive multi-party meetings and conferences with off-the-shelf equipment and conventional Internet connections. We have also confirmed that single-camera volumetric capture systems are acceptable for scenarios that do not require 6DoF, but still side views of users who lack full volume reconstruction may be noticeable. In addition, our study has shown that the adoption of a gamified task, like a guessing game along the lines of charades, has not only resulted very welcome by participants but also very useful to obtain relevant insights into the formulated research questions. This suggests that these types of tasks are appropriate for evaluating communication and interaction technologies, like multi-party holoportation.

Future works will target overcoming the limitations of various technological and methodological aspects, many of which were identified in Section V. First, efforts will be especially devoted to optimizing technological aspects of the end-to-end volumetric media pipeline, specifically in terms of increased resolution and performance, scalability, and adaptability to heterogeneous environments. Adding extra interactive features (e.g., interaction with the environment, collaborative tools, etc.) will also be considered. Second, plans are in the works for further evaluations, including additional test conditions, tasks and use cases. In particular, they will include comparing this platform with other baseline and alternative solutions, such as face-to-face scenarios, traditional conferencing tools, other social VR platforms, and different user representation formats (e.g., avatars, webcams, single-cam point clouds, full volumetric point clouds, and others). Third, we plan to assess the allowable limits on delay and on delay differences (between streams and between clients) in multi-party holoportation scenarios. As done in other works for traditional videoconferencing (e.g., [13, 40]) and social TV (e.g., [3, 40]), these later research questions can be explored by forcing specific network conditions and deploying the service in different scenarios. Fourth, we will additionally assess the visual quality of user representations by means of objective evaluation metrics for point clouds ([52-54]), specifically by considering different capture sensors, encoding settings, network scenarios, and different evaluation metrics in order to correlate each of these aspects with the perceived visual quality.

## ACKNOWLEDGMENT

The authors would like to thank the members of the EU H2020 VR-Together consortium for their valuable contributions, especially Marc Martos and Mohamad Hjeij for their support in developing and evaluating tasks. This work has been partially funded by: the EU's Horizon 2020 program, under agreement nº 762111 (VR-Together project); by ACCIÓ (Generalitat de Catalunya), under agreement COMRDI18-1-0008 (ViVIM project); and by Cisco Research and the Silicon Valley Community Foundation, under the grant Extended Reality Multipoint Control Unit (ID: 1779376). The work by Mario Montagud has been additionally funded by Spain's Agencia Estatal de Investigación under grant RYC2020-030679-I (AEI / 10.13039/501100011033) and by Fondo Social Europeo. The work of David Rincón was supported by Spain's Agencia Estatal de Investigación within the Ministerio de Ciencia e Innovación under Project PID2019-108713RB-C51 MCIN/AEI/10.13039/501100011033

## REFERENCES

[1] S. Firestone, T. Ramalingam, S. Fry, "Voice and Video Conferencing Fundamentals", Cisco Press, 397 pages, March 2007, ISBN-10: 1-58705-268-7.
[2] M. Schmitt, J. Redi, D.C.A. Bulterman, P. Cesar, "Towards individual QoE for multi-party video conferencing", IEEE Transactions on Multimedia (TMM), 20(7):1781-1795, 2018.
[3] F. Boronat, M. Montagud, P. Salvador, J. Pastor, "Wersync: A web platform for synchronized social viewing enabling interaction and collaboration", Journal of Network and Computer Applications, Volume 175, February 2021.
[4] J. Hacker, et al., "Virtually in this together – how web-conferencing systems enabled a new virtual togetherness during the COVID-19 crisis", 29(5), 563-584, European Journal of Information Systems, August 2020.
[5] M. Montagud, J. Li, G. Cernigliaro, A. El Ali, S. Fernandez, P. Cesar, "Towards SocialVR: Evaluating a Novel Technology for Watching Videos Together", Virtual Reality, May 2022 (Online).
[6] J. Jansen, S. Subramanyam , R. Bouqueau, G. Cernigliaro, M. Martos, F. Pérez, P. Cesar, "A Pipeline for Multiparty Volumetric Video Conferencing: Transmission of Point Clouds over Low Latency DASH", ACM Multimedia Systems Conference (MMSys) 2020, Istanbul (Turkey), June 2020.
[7] S. Fernández, M. Montagud, G. Cernigliaro, D. Rincón, "Toward Hyper-realistic and Interactive Social VR Experiences in Live TV Scenarios", IEEE Transactions on Broadcasting, 2021 (Early Access)
[8] N. Khojasteh, A.S. Won, "Working Together on Diverse Tasks: A Longitudinal Study on Individual Workload, Presence and Emotional Recognition in Collaborative Virtual Environments", Frontiers in Virtual Reality, 2:643331, June 2021

[9] M. Chessa, F. Solari, "The sense of being there during online classes: analysis of usability and presence in web-conferencing systems and virtual reality social platforms", Behaviour & Information Technology, 40:12, 1237-1249, 2021.
[10] M. H. Hajiesmaili, L. T. Mak, Z. Wang, C. Wu, M. Chen and A. Khonsari, "Cost-Effective Low-Delay Design for Multiparty Cloud Video Conferencing", IEEE Transactions on Multimedia, vol. 19, no. 12, pp. 2760-2774, Dec. 2017
[11] Y. Wu, C. Wu, B. Li, F.C.M. Lau, "VSkyConf: cloud-assisted multi-party mobile video conferencing", Second ACM SIGCOMM workshop on Mobile cloud computing (MCC '13), 33–38, Hong Kong (China), August 2013.
[12] Y. Lu, Y. Zhao, F. Kuipers, P. Van Mieghem, "Measurement study of multi-party video conferencing", International Conference on Research in Networking, Springer, pp. 96-108, Chennai (India), May 2010.
[13] G. Berndtsson, et al., "Methods for Human-Centered Evaluation of MediaSync in Real-Time Communication". In: M. Montagud, P. Cesar, F. Boronat, J. Jansen J. (eds) "MediaSync: Handbook on Multimedia Synchronization", Springer, 2018.
[14] M. Schmitt, J. Redi, and P. Cesar, "Towards context-aware interactive Quality of Experience evaluation for audiovisual multiparty conferencing," in Proceedings of the International Workshop on Perceptual Quality of Systems, (PQS2016), Berlin, Germany, August 29-31, 2016
[15] S. Loreto, S. P. Romano, "Real-time communication with WebRTC: peer-to-peer in the browser", O'Reilly Media Inc., 164 pages, 2014, ISBN 978-1449371876
[16] M. Wijnants, J. Dierckx, P. Quax, W. Lamotte, "Synchronous MediaSharing: social and communal media consumption for geographically dispersed users", ACM Multimedia Systems Conference (MMSys) 2012, pp. 107-112, Chapel Hill, North Carolina (USA), February 2012
[17] J. Belda, M. Montagud, F. Boronat, M. Martinez, J. Pastor, "Wersync: A web-based platform for distributed media synchronization and social interaction", ACM TVX 2016, Brussels (Belgium), June 2016
[18] S. Rothe, A. Schmidt, M. Montagud, D. Buschek, H. Hußman, "Social viewing in cinematic virtual reality: a design space for social movie applications", Virtual Reality, October 2020
[19] J. Li, et al., "Measuring and Understanding Photo Sharing Experiences in Social Virtual Reality", ACM CHI 2019, Glasgow (UK), May 2019
[20] A. Bleakley, V. Wade, B. R. Cowan, "Finally a Case for Collaborative VR? The Need to Design for Remote Multi-Party Conversations", 2nd ACM Conference on Conversational User Interfaces (CUI '20), Article 25, 1–3, Bilbao (Spain), July 2020
[21] K. Kilteni, R. Groten, M. Slater, "The sense of embodiment in virtual reality", Presence: Teleoperators and Virtual Environments, 21, 4, 373–387, 2012
[22] P. Heidicker, E. Langbehn, F. Steinicke, "Influence of avatar appearance on presence in Social VR", IEEE Symposium on 3D User Interfaces (3DUI) 2017, 233–234, Los Angeles (CA, USA), March 2017.
[23] D. Roth, et al., "Avatar realism and social interaction quality in virtual reality", IEEE Virtual Reality (VR) 2016, 277–278, Greenville (South Carolina, USA), March 2016
[24] H. J. Smith and M. Neff, "Communication behavior in embodied Virtual Reality", ACM Conference on Human Factors in Computing Systems (CHI) 2018, Montreal (QC Canada), April 2018
[25] M. E. Latoschik, et al., "The effect of avatar realism in immersive social virtual realities", 23rd ACM Symposium on Virtual Reality Software and Technology (VRST'17), Gothenburg Sweden November 8 - 10, 2017
[26] T. Waltemate, et al. "The impact of avatar personalization and immersion on virtual body ownership, presence, and emotional response", IEEE transactions on visualization and computer graphics, 24, 4, 1643–1652, 2018
[27] P. Kauff, O. Schreer, "An Immersive 3D Video-Conferencing System Using Shared Virtual Team User Environments", 4th International Conference on Collaborative Virtual Environments (CVE '02), Bonn (Germany), 2002.
[28] R. Mekuria, P. Cesar, I. Doumanis, A. Frisiello, "Objective and subjective quality assessment of geometry compression of reconstructed 3D humans in a 3D virtual room", Proc. SPIE 9599, Applications of Digital Image Processing XXXVIII, Vol. 95991, September 2015
[29] M. McGill, J. H. Williamson, S. Brewster, "Examining the role of smart TVs and VR HMDs in synchronous at-a-distance media consumption", ACM Transactions on Computer-Human Interaction (TOCHI), 23, 5, 33, 2016
[30] S. Gunkel, M. Prins, H. Stokking, O. Niamut, "Social VR Platform: Building 360-degree Shared VR Spaces", ACM TVX 2017, Hilversum (The Netherlands), June 2017

[31] D. S. Alexiadis, A. Chatzitofis, N. Zioulis, O. Zoidi, G. Louizis, D. Zarpalas, P. Daras, "An integrated platform for live 3D human reconstruction and motion capturing", IEEE Transactions on Circuits and Systems for Video Technology, 27, 4, 798–813, 2016

[32] K. Christaki, E. Christakis, P. Drakoulis, A. Doumanoglou, N. Zioulis, D. Zarpalas, P. Daras, "Subjective Visual Quality Assessment of Immersive 3D Media Compressed by Open-Source Static 3D Mesh Codecs", 25th International Conference on Multimedia Modeling (MMM) 2019, Thessaloniki (Greece), January 2019

[33] S. Schwarz et al., "Emerging MPEG Standards for Point Cloud Compression", IEEE Journal on Emerging and Selected Topics in Circuits and Systems, vol. 9, no. 1, pp. 133-148, March 2019

[34] L. Keselman, J. I. Woodfill, A. Grunnet-Jepsen, A. Bhowmik, "Intel realsense stereoscopic depth cameras", IEEE Conference on Computer Vision and Pattern Recognition Workshops, pp. 1-10, 2017

[35] M. Tölgyessy, M. Dekan, Ľ. Chovanec, P. Hubinský, "Evaluation of the Azure Kinect and Its Comparison to Kinect V1 and Kinect V2", Sensors, 21(413), 2021

[36] R. Mekuria, K. Blom, P. Cesar, "Design, Implementation, and Evaluation of a Point Cloud Codec for Tele-Immersive Video", IEEE Transactions on Circuits and Systems for Video Technology", 27(4), 828–842, 2017

[37] M. F. Ursu, et al., "Orchestration: TV-like mixing grammars applied to video-communication for social groups", 21st ACM international conference on Multimedia (ACM MM 2013), 333–342, Barcelona (Spain), October 2013

[38] W. Weiss, R. Kaiser, M. Falelakis, "Orchestration for Group Videoconferencing: An Interactive Demonstrator", 16th ACM International Conference on Multimodal Interaction (ICMI '14), Istanbul (Turkey), 2014

[39] G. J. Sullivan, et al., "New Standardized Extensions of MPEG4-AVC/H.264 for Professional-Quality Video Applications", IEEE International Conference on Image Processing, (ICIP) 2007, San Antonio, Texas (USA), 2007

[40] M. Montagud, P. Cesar, J. Jansen, F. Boronat (Eds.), "MediaSync: Handbook on Multimedia Synchronization" (23 chapters), Springer-Verlag, ISBN 978-3-319-65840-7, 2018

[41] M. Montagud, F. Boronat, H. Stokking, "Early Event-Driven (EED) RTCP Feedback for Rapid IDMS", The 21st ACM International Conference on Multimedia (ACM MM 2013), Barcelona (Spain), October 2013

[42] F. Boronat, D. Marfil, M. Montagud, J. Pastor, "Hybrid Broadcast/Broadband TV Services and Media Synchronization. Demands, Preferences and Expectations of Spanish Consumers", IEEE Transactions on Broadcasting, 64(3), pp. 52-69, August 2017

[43] R. S. Kennedy, N. E. Lane, K. S. Berbaum, Mi. G. Lilienthal, "Simulator Sickness Questionnaire: An Enhanced Method for Quantifying Simulator Sickness", The International Journal of Aviation Psychology, 3:3, 203-220, 1993

[44] System Usability Scale (SUS) Questionnaire https://www.usability.gov/how-to-and-tools/methods/system-usability-scale.html, Last Access in July 2021

[45] NASA Task Load Index (TLX) Questionnaire https://humansystems.arc.nasa.gov/groups/tlx/, Last Access in July 2021.

[46] M. Montagud, J. A. De Rus, R. Fayos-Jordán, M. Garcia-Pineda J. Segura-Garcia, "Open-Source Software Tools for Measuring Resources Consumption and DASH Metrics", ACM MMSYS 2020, Istanbul (Turkey), June 2020

[47] M. Schmitt, J. Redi, P. Cesar, D. Bulterman, "1Mbps is enough: Video quality and individual idiosyncrasies in multiparty HD video-conferencing", Eighth International Conference on Quality of Multimedia Experience (QoMEX), 2016

[48] C. Lee, S. Woo, S. Baek, "Bitrate and Transmission Resolution Determination Based on Perceptual Video Quality", 10th International Conference on Information, Intelligence, Systems and Applications (IISA), 2019

[49] D. R. Thomas, "A general inductive approach for analyzing qualitative evaluation data", American journal of evaluation, 27(2), 237–246, 2006

[50] N. Numan, F. t. Haar, P. Cesar, "Generative RGB-D Face Completion for Head-Mounted Display Removal", IEEE Conference on Virtual Reality and 3D User Interfaces Abstracts and Workshops (VRW) 2021, 2021, pp. 109-116, Virtual Event, April 2021

[51] G. Cernigliaro, M. Martos, M. Montagud, A. Ansari, S. Fernandez, "PC-MCU: point cloud multipoint control unit for multi-user holoconferencing systems", 30th ACM Workshop on Network and Operating Systems Support for Digital Audio and Video (NOSSDAV '20), Istanbul (Turkey), June 2020

[52]   L. A. da Silva Cruz, et al. "Point cloud quality evaluation: Towards a definition for test conditions." IEEE Eleventh International Conference on Quality of Multimedia Experience (QoMEX'19), Berlin (Germany), June 2019.
[53]   I Viola, P. Cesar, "A reduced reference metric for visual quality evaluation of point cloud contents" IEEE Signal Processing Letters 27, 1660-1664, 2020.
[54]   D. Lazzarotto, E. Alexiou, T. Ebrahimi, "Benchmarking of objective quality metrics for point cloud compression", IEEE 23rd International Workshop on Multimedia Signal Processing (MMSP'21), Tampere (Finland), October 2021